\documentclass[showpacs,prl,aps,twocolumn]{revtex4}

 \usepackage{bm}
 \usepackage{amsmath}
 \usepackage{amssymb}
 \usepackage{latexsym} 
 \usepackage{amsfonts} 
 \usepackage{epsfig}
 \usepackage{psfrag} 

 \newcommand{\ket}[1]{\left|#1\right>}

 \newcommand{\ul}{\underline} 
 \newcommand{\f}[1]{\mbox{\boldmath$#1$}}
 \newcommand{\fk}[1]{\mbox{\boldmath$\scriptstyle#1$}}
 
 \newcommand{\na}{\mbox{\boldmath$\nabla$}}
 \newcommand{\bea}{\begin{eqnarray}}
 \newcommand{\ea}{\end{eqnarray}}
 \newcommand{\eea}{\end{eqnarray}}
 \newcommand{\ord}{{\cal O}}

\begin{document}
  
\title{Signatures of the Unruh effect from electrons accelerated by
  ultra-strong laser fields}  

\author{Ralf Sch\"utzhold and Gernot Schaller}

\affiliation{Institut f\"ur Theoretische Physik, 
Technische Universit\"at Dresden, D-01062 Dresden, Germany}


\author{Dietrich Habs}

\affiliation{
Department f\"ur Physik der Ludwig-Maximilians-Universit\"at M\"unchen
und Maier-Leibnitz-Laboratorium, 
Am Coulombwall 1, D-85748 Garching, Germany}

\begin{abstract}
We calculate the radiation resulting from the Unruh effect for
strongly accelerated electrons and show that the photons are created
in pairs whose polarizations are maximally entangled. 
Apart from the photon statistics, this quantum radiation can further
be discriminated from the classical (Larmor) radiation via the
different spectral and angular distributions. 
The signatures of the Unruh effect become significant if the external 
electromagnetic field accelerating the electrons is not too
far below the Schwinger limit and might be observable with future
facilities.  
Finally, the corrections due to the birefringent nature of the QED
vacuum at such ultra-high fields are discussed. 
\end{abstract}


\maketitle

{\em Introduction}\quad
%
One of the most fascinating phenomena of non-inertial quantum field
theory is the Unruh effect:
An observer or detector undergoing a uniform acceleration $a$
experiences the Minkowski vacuum as a thermal bath with the Unruh
temperature \cite{unruh}
\bea
\label{unruh}
T_{\rm Unruh}
=
\frac{\hbar}{2\pi k_{\rm B}c}\,a
\,.
\ea
As one might expect from the principle of equivalence, the Unruh
effect is closely related to Hawking radiation \cite{hawking}, 
i.e., black hole evaporation: 
The uniformly accelerated observer (detecting the Unruh effect) 
corresponds to an observer at a fixed distance to the horizon 
(feeling the gravitational pull and measuring the Hawking radiation);
whereas the inertial observer in flat space-time is analogous to an
unfortunate astronaut freely falling into the black hole.
However, there is also a crucial difference between the two
phenomena: 
In contrast to the case of uniform acceleration, the free fall into a
black hole is (per definition of a black hole) not invariant under
time-reversal. 
Hence, while Hawking radiation generates a real out-flow of energy 
(black hole evaporation), the Unruh effect corresponds to an
equilibrium thermal bath and does not create any energy flux 
{\em per se}. 

The most direct way of observing this striking effect would be to
accelerate a detector and to measure its excitations.
However, this is extremely difficult since moderate accelerations
correspond to extremely low temperatures and thus, the Unruh effect
has not been directly observed so far  
(see, however, \cite{orbit,rosu}). 
Therefore, we focus on a somewhat indirect signature in the following: 
Since the uniformly accelerated detector acts as if it was immersed in
a thermal bath, there is a finite probability that it absorbs a virtual 
particle from this bath and passes to an excited state.
Translated back into the inertial frame, this process corresponds to
the emission of a real particle \cite{happen}.
The opposite process, when the detector re-emits the virtual particle
into the bath in the accelerated frame and goes back to its ground
state, also corresponds to the emission of a real particle in the
inertial frame.

In the limiting case that the time between absorption and re-emission
becomes arbitrarily small, the detector transforms into a scatterer
which scatters (virtual) particles from one mode into another mode of
the thermal bath in the accelerated frame.
Translated back into the inertial frame, this process corresponds to
the emission of two real particles by the accelerated scatterer.
This effect is analogous to moving-mirror radiation \cite{birrell}
and can be interpreted as a signature of the Unruh effect.
In the following, we calculate this quantum radiation given off
by point-like non-inertial scatterers (as a model for electrons
accelerated in ultra-intense electromagnetic fields) and compare it to 
the classical (Larmor) radiation.
An analogous idea has already been pursued in \cite{chen} but in the 
derivation presented therein did not take into account crucial
features of the radiation (such as the fact that the photons are
always created in correlated pairs). 

{\em The Model}\quad
%
Assuming that the electric field $\f{E}$ is much stronger than the
magnetic field $\f{B}$ in the region of interest $\f{E}^2\gg c^2\f{B}^2$,
we neglect the spin of the electron since the related energy 
$\f{\mu}_e\cdot\f{B}$ is much smaller than the other relevant
energies (such as the Unruh temperature). 
For electromagnetic waves whose wavelength is much larger than the 
Compton wavelength, the electron acts as a classical point-like
scatterer with an energy-independent cross section (Thomson scattering).
Since we are mostly interested in backward scattering 
(reflection, see below), we further neglect the angular dependence
of the scattering amplitude ($s$-wave scattering approximation).
Under these assumptions, the electromagnetic field under the influence
of such a point-like scatterer (i.e., electron) with the trajectory
$\f{r}_e[t]$ is governed by the Lagrangian density
($\hbar=\varepsilon_0=\mu_0=c=1$ throughout) 
\bea
\label{ansatz}
{\cal L}
=
\frac12\left(\f{E}^2-\f{B}^2\right)
-
g\f{A}^2\delta^3(\f{r}_e[t]-\f{r})\sqrt{1-\f{\dot r}_e^2[t]}
\,,
\ea
where the coupling $g$ determines the $s$-wave scattering length and
$\f{A}$ is the vector potential in radiation gauge.
The last factor ensures the relativistic invariance of the action 
${\cal A}=g\int ds\,A_\mu A^\mu-\int d^4x\,F_{\mu\nu}F^{\mu\nu}/4$.  

This model can be motivated by the following simple (non-relativistic)
picture:
The charge $q$ of the electron determines its density 
$\varrho(t,\f{r})=q\delta^3(\f{r}_e[t]-\f{r})$ 
and its current $\f{j}=\f{\dot r}_e\varrho$. 
Omitting the magnetic field, we get 
$\f{\ddot r}_e=q\f{E}/m=q\f{\dot A}/m$, and neglecting the spatial
dependence (due to $\f{\dot r}_e\perp\f{k}$) of $\f{A}$, 
this can be solved via $\f{\dot r}_e=q\f{A}/m$.
Insertion of the resulting current
$\f{j}=q^2\delta^3(\f{r}_e[t]-\f{r})\f{A}/m$
into the Lagrangian $\f{j}\cdot\f{A}$ reproduces our 
ansatz~(\ref{ansatz}) with $g=q^2/m$ which yields the correct cross
section for planar Thomson scattering.
In the natural units used here, the charge $q$ is related to the 
fine-structure constant via $q=\sqrt{4\pi\alpha_{\rm QED}}\approx0.3$.  

{\em Particle Creation}\quad
%
In order to calculate the photons created by the non-inertial motion
$\f{r}_e[t]$ of the scatterer, let us split the total Hamiltonian into
a perturbation part 
\bea
\label{Hamiltonian}
\hat H_1(t)
=
g\f{\hat A}^2(t,\f{r}_e[t])\sqrt{1-\f{\dot r}_e^2[t]}
\,,
\ea
supplemented with the usual adiabatic switching on and off 
$g(|t|\uparrow\infty)=0$, and the undisturbed Hamiltonian 
$\hat H_0=\frac12\int d^3r(\f{\hat E}^2+\f{\hat B}^2)$, 
which leads to the usual normal mode expansion.

Since the coupling $g$ is much smaller than the other relevant length
scales (such as the wavelengths of the photons), the evolution of the
initial Minkowski vacuum $\ket{0}$ can be derived via time-dependent
perturbation theory  
\bea
\label{perturbation}
\ket{{\rm out}}
&=&
\ket{0}-i\int dt\,\hat H_1(t)\ket{0}
+\ord(g^2)
\\
&=& 
\nonumber
\ket{0}+\sum\limits_{\fk{k},\lambda,\fk{k'},\lambda'}
{\mathfrak A}_{\fk{k},\lambda,\fk{k'},\lambda'}
\ket{\f{k},\lambda,\f{k'},\lambda'}
+\ord(g^2)
\,,
\ea
with the the two-photon amplitude 
\bea
\label{two-photon}
{\mathfrak A}_{\fk{k},\lambda,\fk{k'},\lambda'}
&=& 
\frac{\f{e}_{\fk{k},\lambda}\cdot\f{e}_{\fk{k'},\lambda'}}{2iV\sqrt{kk'}}
\int dt\,
g\,\sqrt{1-\f{\dot r}_e^2[t]}\;
\times
\\
&&
\nonumber
\times
\exp\left\{i(k+k')t-i(\f{k}+\f{k'})\cdot\f{r}_e[t]\right\}
\,.
\ea
Here $\f{k}$ is the wavenumber and $k=|\f{k}|$ the frequency of the
photon modes; $\lambda$ labels their polarization described by the
unit vector $\f{e}_{\fk{k},\lambda}$; and $V$ is the quantization
volume. 
Since a time-resolved detection of the created photons is probably
infeasible, polarization and momentum are the best observables to be
measured. 
As one may infer from the above expression, the photons are always
emitted in pairs (squeezed state) and there is a perfect entanglement
of the polarizations of the two photons due to the scalar product 
$\f{e}_{\fk{k},\lambda}\cdot\f{e}_{\fk{k'},\lambda'}$; e.g., 
parallel photons $\f{k}\parallel\f{k'}$ must have the same
polarization $\lambda=\lambda'$.
Note that this applies to linear polarization, the circular
polarizations of the two created photons are opposite 
(for Thomson scattering) due to angular momentum conservation. 

In terms of the new integration variable 
$\tau=t-r^\|_e[t]$ with $r^\|_e=(\f{k}+\f{k'})\cdot\f{r}_e/(k+k')$,
the two-photon amplitude, 
\bea
\label{new-Unruh}
{\mathfrak A}_{\fk{k},\lambda,\fk{k'},\lambda'}
=
\frac{\f{e}_{\fk{k},\lambda}\cdot\f{e}_{\fk{k'},\lambda'}}{2iV\sqrt{kk'}}
\int d\tau\,g\,
\frac{\sqrt{1-\f{\dot r}_e^2[t]}}{1-\dot r^\|_e[t]}
e^{i(k+k')\tau}
\,,
\ea
is determined by the Fourier transform of the effective
(direction-dependent) Doppler factor in the integral above.  

{\em Larmor Radiation}\quad
%
In order to discuss the observability of this quantum radiation, it
must be compared with the competing classical process.
The Larmor radiation can be derived from the relativistic action
$q\int dx^\mu\,A_\mu$ and corresponds to a coherent state 
\bea
\label{coherent}
\ket{\alpha}
=
\prod\limits_{\fk{k},\lambda}\exp\left\{
\alpha_{\fk{k},\lambda}\hat a^\dagger_{\fk{k},\lambda}-
\alpha^*_{\fk{k},\lambda}\hat a_{\fk{k},\lambda}
\right\}
\ket{0}
\,,
\ea
with the coefficients (see, e.g., \cite{scully})
\bea
\label{coefficients}
\alpha_{\fk{k},\lambda}
=
q\int dt\,
\frac{\f{e}_{\fk{k},\lambda}\cdot\f{\dot r}_e[t]}{\sqrt{2Vk}}\,
e^{ikt-i\fk{k}\cdot\fk{r}_e[t]}
\,.
\ea
The numerator $\f{e}_{\fk{k},\lambda}\cdot\f{\dot r}_e$ displays the
well-known blind spot in forward and backward direction
$\f{k}\|\f{\dot r}_e$. 
The introduction of a new integration variable 
$\tau=t-\f{k}\cdot\f{r}_e[t]/k$ 
\bea
\label{new-Larmor}
\alpha_{\fk{k},\lambda}
=
q\int d\tau\,
\frac{\f{e}_{\fk{k},\lambda}\cdot\f{\dot r}_e}{1-\f{e}_k\cdot\f{\dot r}_e}
\,
\frac{e^{ik\tau}}{\sqrt{2Vk}}
\,,
\ea
yields a Fourier transform similar to Eq.~(\ref{new-Unruh}).

For an investigation of the detectability of the quantum radiation in
Eq.~(\ref{new-Unruh}), the two-photon amplitude 
$|{\mathfrak A}_{\fk{k},\lambda,\fk{k'},\lambda'}|$ must be compared
with the amplitude for the competing classical process, which is given
by $|\alpha_{\fk{k},\lambda}\alpha_{\fk{k'},\lambda'}|$ to lowest
order in $\alpha_{\fk{k},\lambda}$.
In view of the smallness of the coupling $g$ and assuming comparable
results of the Fourier integrals (no resonances etc.), there are
basically two possibilities for achieving 
$|{\mathfrak A}_{\fk{k},\lambda,\fk{k'},\lambda'}|>
|\alpha_{\fk{k},\lambda}\alpha_{\fk{k'},\lambda'}|$: 
small velocities $\f{\dot r}_e^2\ll1$ or small angles~$\vartheta$
between $\f{k}$ and $\f{\dot r}_e$ (blind spot).
The first alternative is probably impractical since the total effect
becomes too small, but the latter option can be realized for a
uni-directional acceleration.
For typical $k$-values, quantum radiation becomes comparable to
classical radiation for
\bea
\label{max}
\vartheta^2_{\rm max}
=
\ord\left(\frac{gk}{q^2}\right)
\,,
\ea
i.e., inside a small forward and backward cone (blind spot).
The total probability of emitting a photon pair inside these cones of
``quantum domination'' scales as 
\bea
\label{total}
{\mathfrak P}(\vartheta_{\rm max})
=
\sum\limits_{\fk{k},\lambda,\fk{k'},\lambda'}^{\vartheta<\vartheta_{\rm max}}
|{\mathfrak A}_{\fk{k},\lambda,\fk{k'},\lambda'}|^2
=
\ord\left(\frac{g^4k^4}{q^4}\right)
\,.
\ea
The typical $k$-value is set by the characteristic time-scale of the
trajectory. 
For a smooth (e.g., Gaussian) electric field pulse $\f{E}(t)$ which
accelerates the electron to moderately relativistic velocities (say
$\gamma=2$), this would be the pulse length.
Assuming a rather short pulse of order 0.1 attoseconds, i.e., 
$\Delta t=\ord(10^{-19}\,{\rm s})$, comparison with the coupling 
$g\approx3.5\times10^{-14}\,{\rm m}$ yields $gk/q^2=\ord(10^{-2})$,
i.e., quantum radiation dominates within a forward/backward cone of a
few degrees.  
Unfortunately, the total probability ${\mathfrak P}(\vartheta_{\rm max})$
of emitting a photon pair in these cones is suppressed by several
orders of magnitude and hence probably very hard to detect.

In order to measure the quantum radiation, far shorter characteristic 
time-scales are desirable.
One option could be a pulse with a sharp front end whose typical
raising time $\delta t\ll\Delta t$ is much shorter than the pulse
length $\Delta t$.
However, even for a rectangular electric field pulse $\f{E}(t)$, the 
Fourier transform of $\f{\dot r}_e$ decays like $1/(k+k')^2$ for large
wavenumbers. 
Hence, the enhancement of ${\mathfrak P}(\vartheta_{\rm max})$ would
merely be logarithmic $\ln^2(\Delta t/\delta t)$ for non-relativistic or
moderately relativistic velocities -- which is probably not sufficient.
Unless further amplification mechanisms such as resonances are
present, the only possibility left 
(apart from generating much shorter pulses, which is extremely hard) 
is to use long pulses with strong electric fields which accelerate the
electrons to ultra-relativistic velocities and to exploit the large
Lorentz boost factors $\gamma\gg1$. 

{\em Ultra-relativistic Regime}\quad
%
As motivated above, let us now consider ultra-relativistic velocities
$\gamma\gg1$. 
Since both, quantum and classical radiation will be boosted forward in
this case, we shall focus on a small forward cone
$\vartheta\ll1/\gamma$, see also Eq.~(\ref{angle-rel}) below.
In this limit, the integral~(\ref{new-Unruh}) simplifies to 
(for $\lambda=\lambda'$)
\bea
\label{Unruh-rel}
{\mathfrak A}_{\fk{k},\lambda,\fk{k'},\lambda'}
\approx
\int d\tau\,\frac{g\gamma(\tau)}{iV\sqrt{kk'}}\,
e^{i(k+k')\tau}
\,,
\ea
with a time-dependent Lorentz factor $\gamma(\tau)$ whose evolution is
determined by $d\gamma/dt\approx qE(t)/m$ as well as 
$dt\approx2\gamma^2d\tau$.
The leading contribution of the above integral arises near the
maximum Lorentz factor $\gamma_{\rm max}$, i.e., in the final stage of 
the acceleration phase, where $d\gamma/d\tau$ drops from 
$2\gamma^2qE/m$ to zero on an effective time scale of 
$\Delta\tau=\ord(\Delta t/\gamma^2_{\rm max})$, which determines the
cutoff wavenumber via $k_{\rm cut}=\ord(\gamma^2_{\rm max}/\Delta t)$.
Since $\gamma_{\rm max}$ is roughly given by the pulse length 
$\Delta t$ times the acceleration $qE/m$, the cutoff wavenumber 
can alternatively be written as 
$k_{\rm cut}=\ord(\gamma_{\rm max}qE/m)$.
Below this cutoff, the above integral behaves like the Fourier
transform of a Heaviside step function of height $\gamma_{\rm max}$
\bea
\label{Unruh-leading}
{\mathfrak A}_{\fk{k},\lambda,\fk{k'},\lambda'}
=
\ord\left(
\frac{g\gamma_{\rm max}}{V\sqrt{kk'}(k+k')}
\right) 
\,.
\ea
A similar estimate for the Larmor radiation yields 
\bea
\label{Larmor-rel}
\alpha_{\fk{k},\lambda}
\approx
\int d\tau\,
\frac{2q\vartheta\gamma^2}{\sqrt{2Vk}}\,
e^{ik\tau}
=
\ord\left(
\frac{q\vartheta\gamma^2_{\rm max}}{\sqrt{Vk^3}}
\right) 
\,,
\ea
with basically the same wavenumber cutoff.
Of course, quantum radiation again dominates for sufficiently
small~$\vartheta$.
For the cutoff wavenumber $k_{\rm cut}$, the size of the small forward
cone of ``quantum domination'' scales as 
\bea
\label{angle-rel}
\vartheta_{\rm max}
=\ord\left(
\sqrt{\frac{qE}{m^2}}\,\frac1{\gamma_{\rm max}}\right) 
\,,
\ea
and is determined by $\gamma_{\rm max}$ and ratio of the 
electric field $E$ over the Schwinger \cite{schwinger} limit
$E_S=m^2/q$. 
The probability of (two-photon) emission in this cone is given by 
\bea
\label{probability-rel}
{\mathfrak P}(\vartheta_{\rm max})
=
\sum\limits_{\fk{k},\lambda,\fk{k'},\lambda'}^{\vartheta<\vartheta_{\rm max}}
|{\mathfrak A}_{\fk{k},\lambda,\fk{k'},\lambda'}^2|
=
\ord\left(q^4\,\frac{E^4}{E_S^4}\right)
\,.
\ea
Interestingly, for a given electric field strength $E$, this
probability does not depend significantly on the pulse length since  
$\gamma_{\rm max}$ cancels -- but the energy $k_{\rm max}$ and the
angular distribution $\vartheta_{\rm max}$ of the emitted photons
does. 

{\em Extensions}\quad
%
Since the observation of the photon pairs requires electromagnetic
fields which are not too far below the Schwinger limit, one should
also consider the impact of these ultra-high fields on the QED vacuum,
which then acts as a medium and displays effects such as
bi\-refringence.  
To this end, we consider the first non-linear corrections from the
Euler-Heisenberg Lagrangian \cite{euler}
\bea
{\cal L}
=
\frac12\left(\f{E}^2-\f{B}^2\right)+
\frac{(\f{E}^2-\f{B}^2)^2+7(\f{E}\cdot\f{B})^2}{90\pi E_S^2/\alpha_{\rm QED}}
\,.
\ea
In media, the dielectric displacement 
$\f{D}=\partial{\cal L}/\partial\f{E}$ may differ from the electric
field $\f{E}$ and hence the first Maxwell equation $\na\cdot\f{D}=0$
does not necessarily imply ${\na\cdot\f{E}=0}$.
If we neglect the external magnetic field and linearize around an
approximately homogeneous external electric field~$\f{E}_0$, we get 
$\f{D}=\f{\ul\varepsilon}\cdot\f{E}=\f{\ul\varepsilon}\cdot\f{\dot A}$
with the permittivity tensor~$\f{\ul\varepsilon}$.
Hence, the transversality condition 
$\f{k}\cdot\f{\ul\varepsilon}\cdot\f{e}_{\fk{k},\lambda}=0$ 
deviates from 
$\f{k}\cdot\f{e}_{\fk{k},\lambda}=0$ and thus the Larmor
radiation does not necessarily vanish in forward direction anymore. 
However, for a trajectory along a field line $\f{\dot r}_e\|\f{E}_0$,
i.e., along an eigenvector of $\f{\ul\varepsilon}$, the Larmor radiation
still has the blind spot $\f{k}\perp\f{e}_{\fk{k},\lambda}$ 
in forward direction $\f{k}\|\f{E}_0$.

With an external magnetic field $\f{B}_0$, on the other hand,
additional terms appear.
Assuming $\f{k}\|\f{E}_0\perp\f{B}_0$, one of the (linear) polarizations
$\lambda$ has blind spot in forward direction 
$\f{k}\cdot\f{e}_{\fk{k},\lambda}=0$
but the other one $\lambda'$ has not:
\bea
\f{k}\cdot\f{e}_{\fk{k},\lambda'}=
\frac{4\alpha_{\rm QED}}{45\pi}\,
\frac{kE_0B_0}{E_S^2}
\neq0
\,.
\ea
Although the numerical pre-factor is rather small ${2\times10^{-4}}$,
this effect should be taken into account when searching for quantum
radiation \cite{threshold}. 
On the other hand, it might also provide an opportunity for testing
the birefringent nature of the QED vacuum in the presence of
ultra-high external fields (see also \cite{wipf}). 
Similarly, one obtains small corrections to the two-photon
correlations.  

\begin{figure}[ht]
\includegraphics[height=2.1cm]{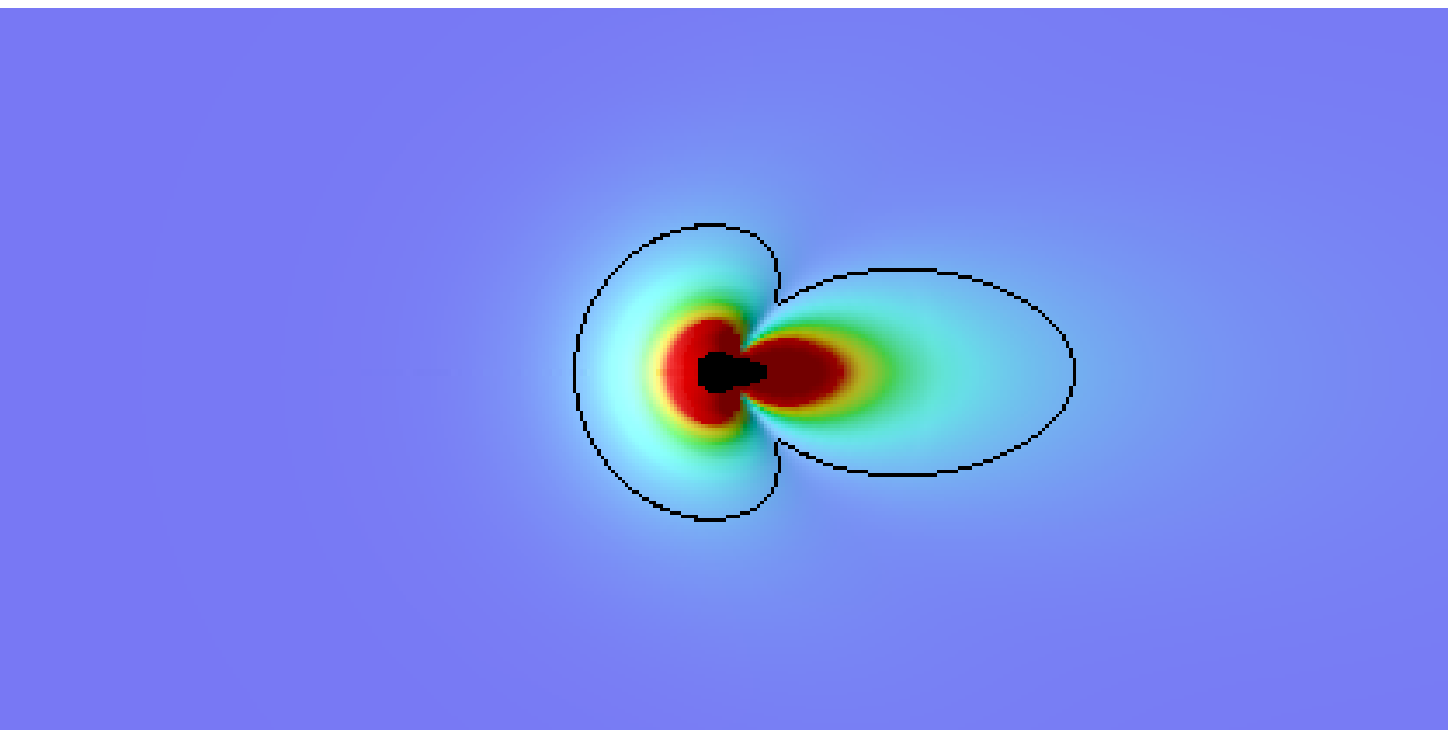}
\includegraphics[height=2.1cm]{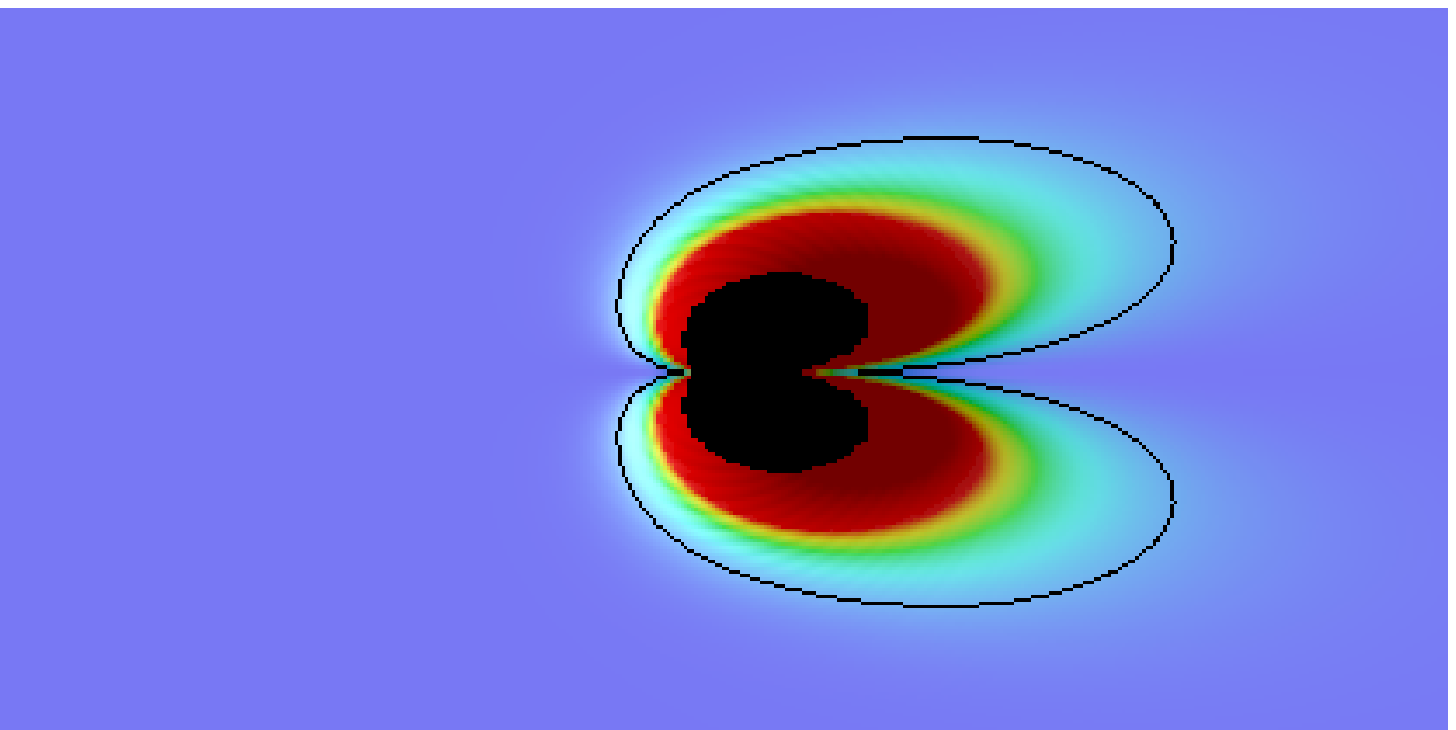}
\caption{\label{figure1} [Color online] Two-photon amplitudes for
  quantum (Unruh, left image) and classical (Larmor, right image)
  radiation for $\f{k}=\f{k'}$ and plotted as a function of
  $\f{k}$. The electron (being initially at rest) is accelerated by a
  Gaussian electric field pulse with a width of 0.3 attoseconds to a
  moderately relativistic velocity $\gamma_{\rm max}\approx2$ and
  moves to the right. The points in the middle of the images
  correspond to $\f{k}=0$ and the maximum $k$-values at the left and
  right boundaries to 30~keV. The color coding (same in both images)
  is chosen such that red  
  indicates large amplitude and blue vanishing amplitude. The black
  areas are the excised singularities at $\f{k}=0$ and the 
  black lines are iso-lines. One can clearly see that quantum
  radiation dominates inside a small forward and backward cone.} 
\end{figure}

{\em Summary}\quad
%
We have studied the conversion of (virtual) quantum vacuum
fluctuations into (real) particles by non-inertial scattering for the
example of strongly accelerated electrons.
This quantum radiation can be discriminated from classical (Larmor)
radiation via the different angular and spectral distributions and the 
distinct photon statistics:
In the quantum case, the photons are always emitted in pairs 
(squeezed state) with maximally entangled polarizations -- whereas
the classical radiation corresponds to a coherent state 
(independent photons with Poissonian statistics).

The probability of emitting two photons with wave\-numbers 
$k\ll k_{\rm cut}=\ord(\gamma_{\rm max}qE/m)$ scales as $1/k^4$ for
quantum (Unruh) radiation and as $1/k^6$ for Larmor.  
For $k$-values larger than this cutoff~$k_{\rm cut}$, 
the decline is faster than polynomial [smooth $C^\infty$-pulse $E(t)$].
Hence the typical photon energies are determined by that cutoff 
$k_{\rm cut}=\ord(\gamma_{\rm max}qE/m)$ which corresponds to the
Unruh temperature (\ref{unruh}) boosted by $\gamma_{\rm max}$ and may
even exceed the electron mass for large Lorentz factors  
(thus enabling us to distinguish this quantum radiation 
from other effects such as annihilation etc.). 
Note, however, that the spectrum is not Planckian in general.
This is caused by the non-trivial translation from the accelerated
frame to the inertial frame with a time-dependent Lorentz boost factor
$\gamma(t)$ whose rate of change is of the same order as the frequency
corresponding to the Unruh temperature (i.e., non-adiabatic). 
Of course, a state consisting of pairs of correlated photons can never
be exactly thermal.

The Larmor radiation has a blind spot in forward (and backward)
direction where the quantum (Unruh) radiation is maximal leading to a
cone of ``quantum domination'' with a small angle 
$\vartheta_{\rm max}$ in Eq.~(\ref{angle-rel}), see also
Fig.~\ref{figure1}. 
In this cone, the probability for emitting a photon pair given by 
Eq.~(\ref{probability-rel}) scales with
the square of the fine-structure constant times the ratio of the
electric field over the Schwinger limit $E_S=m^2/q$ to the fourth
power: 
$(4\pi\alpha_{\rm QED})^2E^4/E_S^4\approx8\times10^{-3}(E/E_S)^4$.
Hence the signatures of the Unruh effect might be detectable in future
facilities (cf.~\cite{mourou,gordienko,bulanov}) generating 
electric fields not too far below the Schwinger limit 
$E_S\approx1.3\times10^{18}\,{\rm V/m}$
(corresponding to an intensity of order $10^{29}\,{\rm W/cm^2}$) 
which accelerate the electrons to ultra-relativistic velocities.
(Conversely, one might use the spectral and angular distribution of
the radiation to determine the characteristic parameters of the pulse
such as $E$, $\Delta t$, and $\gamma_{\rm max}$.) 
The smallness of the above probability could be compensated by 
accelerating many electrons simultaneously or subsequently.
Note that the detection probably also requires coincidence measurements
(exploiting the photon statistics) since the probability for a single
Larmor photon in this cone scales as $q^2E^2/E_S^2$.

\acknowledgments
%
{\em Acknowledgments}\quad
%
R.~S.~acknowledges valuable discussions with Bill Unruh. 
R.~S.~and G.~S.~were supported by the Emmy-Noether Programme of the
German Research Foundation (DFG, \# SCHU~1557/1-2).



\begin{thebibliography}{499}

\bibitem{unruh}
W.~G.~Unruh,
Phys.\ Rev.\ D {\bf 14}, 870 (1976).

\bibitem{hawking}
S.~W.~Hawking,
Nature {\bf 248}, 30 (1974);
Comm.\ Math.\ Phys.\ {\bf 43}, 199 (1975).

\bibitem{orbit}
J.~S.~Bell and J.~M.~Leinaas,
  Nucl.\ Phys.\ B {\bf 284}, 488 (1987);
%
W.~G.~Unruh,
Phys.\ Rept.\ {\bf 307}, 163 (1998).

\bibitem{rosu}
H.~C.~Rosu,
Grav.\ Cosmol.\  {\bf 7}, 1 (2001);
Int.\ J.\ Mod.\ Phys.\ D {\bf 3}, 545 (1994);
Physics World, October 1999, 21-22.

\bibitem{happen}
W.~G.~Unruh and R.~M.~Wald,
Phys.\ Rev.\ D {\bf 29}, 1047 (1984).

\bibitem{birrell}
N.~D.~Birrell and P.~C.~W.~Davies,
{\em Quantum Fields in Curved Space},
(Cambridge University Press, Cambridge, England 1982).

\bibitem{chen}
P.~Chen and T.~Tajima,
Phys.\ Rev.\ Lett.\ {\bf 83}, 256 (1999).

\bibitem{scully}
M.~O.~Scully and M.~S.~Zubairy,
{\em Quantum Optics},
(Cambridge University Press, Cambridge, England 1997).

\bibitem{schwinger}
J.~Schwinger,
Phys.\ Rev.\ {\bf 82}, 664 (1951).

\bibitem{euler}
W.~Heisenberg and H.~Euler, Zeit.\ Phys.\ {\bf 98}, 714 (1936).

\bibitem{threshold}
Of course, the ultra-high electromagnetic fields under consideration
also diminish the threshold for electron-positron pair creation 
(Schwinger effect \cite{schwinger}) leading to a very complicated
structure of the QED  vacuum.
%
For example, the thermal bath experienced by the accelerated observer
would not only contain photons but also electron-positron pairs.
%
These virtual pairs lead to many new and challenging effects. 
%
However, such non-perturbative contributions are typically 
exponentially suppressed $\exp\{-\pi E_S/E\}$ and hence one would
expect that they can be neglected in comparison to polynomially small
effects such as $(E/E_S)^4$ in Eq.~(\ref{probability-rel}) 
sufficiently below the Schwinger limit.

\bibitem{wipf}
T.~Heinzl {\em et al},
{\tt hep-ph/0601076}. 

\bibitem{mourou}
T.~Tajima and G.~Mourou,
Phys.\ Rev.\ ST Accel.\ Beams {\bf 5}, 031301 (2002);
%
G.~Mourou, Phys.\ Today {\bf 51}, 22 (1998).

\bibitem{gordienko}
S.~Gordienko, A.~Pukhov, O.~Shorokhov, and T.~Baeva,
Phys.\ Rev.\ Lett.\ {\bf 94}, 103903 (2005).

\bibitem{bulanov}
S.~V.~Bulanov, T.~Esirkepov, and T.~Tajima, 
Phys.\ Rev.\ Lett.\ {\bf 91}, 085001 (2003).

\end{thebibliography}
\end{document}